%% file: 4node_wsm.tex
\documentclass[aps, prb, reprint]{revtex4-1}
\usepackage{amsmath, amssymb, graphicx, color, mathrsfs, hyperref, array, bbm}

\input{defcommands.tex}

\newcommand{\hlts}{\eta}
\newcommand{\trev}{\mathcal{T}} 
\newcommand{\inv}{\mathcal{I}} 
\newcommand{\cc}{\mathcal{C}} 

\begin{document}
\title{Connecting the dots: \\ Time-reversal symmetric Weyl Semimetals with tunable Fermi arcs}

\author{Vatsal Dwivedi and Srinidhi T Ramamurthy}
\affiliation{Department of Physics and Institute for Condensed Matter Theory, \\ University of Illinois at Urbana-Champaign, IL 61801, USA}

\begin{abstract}
We propose a one-parameter family of noninteracting lattice models for Weyl semimetals with 4 Weyl nodes and tunable Fermi arcs. These 2-band model Hamiltonians are time-reversal symmetric with $\mathrm{T}^2 = + 1$, and tuning the parameter changes the connectivity of the Fermi arcs continuously without affecting the location and chiralities of the Weyl nodes in the bulk Brillouin zone. The bulk polarization and magnetization are shown to vary with this parameter, a dependence inaccessible to the low energy effective field theory. 
\end{abstract}

\maketitle


\section{Introduction}
Weyl semimetals(WSMs), a class of 3+1 dimensional semimetallic phases, are termed ``topological'' because of the existence of isolated Weyl nodes in the bulk Brillouin zone, which cannot be gapped without annihilating them in pairs, either by breaking the translation invariance or by adding interactions\cite{turner-ashvin2011, witten_wsm}. Theoretically proposed about 5 years ago\cite{turner-ashvin2011}, these phases have experienced a resurgance of interest owing to their recent realization in tantalum arsenide (TaAs)\cite{hasan_wsm, hasan_wsm2, hasan_wsm3} and in photonic crystals\cite{marin_wsm_photonic,marin_wsm_photonic2}. 

The existence of WSMs in 3+1 spacetime dimensions can be understood by the interplay of time-reversal($\mathrm{T}$) and inversion($\mathrm{I}$) symmetries. If a Hamiltonian is symmetric under both of these operations, we generically get line nodes\cite{bhbook,hughes-srinidhi_emres}, which are closed loops in the Brillouin zone along which the bulk gap closes and hence the Berry curvature is singular. Breaking either of these symmetries leads to opening of the gap at all but a finite number of points, thereby leading to WSMs\cite{turner-ashvin2011}. 

Given a Weyl node of positive chirality at lattice momentum $\vk$, inversion (time-reversal) dictates that there is another Weyl node at $\vk' = -\vk$ with negative (positive) chirality. For fermions on a lattice, the fermion doubling theorem\cite{nielsen-ninomiya} demands that there be an equal number of positive/negative chirality Weyl nodes. Thus, an inversion symmetric WSM (broken $\mathrm{T}$) has $2n$ Weyl nodes, while a time reversal symmetric WSM must have at least $4n$ nodes, where $n$ is a positive integer. 

The WSMs exhibit nontrivial (\emph{quasi-topological}) electromagnetic(EM) response\cite{hughes-srinidhi_emres,hosur2013}, notable features of which are the anomalous Hall effect\cite{chen2013}(AHE), the chiral magnetic effect(CME) and a realization of the chiral (Adler-Bell-Jackiw) anomaly\cite{zyuzin2012,sid2013}. The discrete symmetries also impose strong constraints on the corresponding transport coefficients. For instance, the AHE requires broken time-reversal symmetry, while the CME requires that both time-reversal and inversion be broken\cite{hughes-srinidhi_emres}.

A remarkable feature of topological insulators is the existence of nontrivial surface states which are attached to the bulk spectrum. These surface states carry information about the bulk band topology, and cannot be gapped out without closing the bulk gap. Similarly, WSMs exhibit \emph{Fermi arcs}, a continuous curve of gapless modes connecting the Weyl nodes of opposite chirality (projected onto the surface Brillouin zone). But for a Weyl semimetal containing more than one pair of nodes, there would be many possible ways to connect them. A natural question to ask is whether these different connectivities can be continuously deformed into each other by tuning only the bulk while maintaining a fixed (Dirichlet, for instance) boundary condition at the surfaces (i.e, not adding any extra surface terms) and without closing any additional gaps in the bulk. 

In this paper, we answer this question in the affirmative, by explicitly constructing a one-parameter family of 2-band tight binding models with 4 Weyl nodes. The location and chiralities of the Weyl nodes in the bulk Brillouin zone are independent of the tuning parameter; however, by varying it one can \emph{rewire} (i.e, change the connectivity of) the Fermi arcs continuously. 

Our models respects time-reversal symmetry (with $\mathrm{T}^2 = +1$), so that both the AHE and the CME coefficients vanish. However, using analytic calculations as well as exact diagonalizations(ED), we show that the tuning of the Fermi arcs manifests itself in a nonzero polarization/magnetization response, which varies continuously with the tuning parameter.
These effects are completely invisible to a linear order low energy effective theory, which only sees the locations and chiralities of the Weyl nodes. We use the transfer matrix formalism discussed by one of us in Ref \onlinecite{vd-vc_tm} for some of the analytic computations on the model. 

The rest of this paper is organized as follows: In \S \ref{sec:wsm} the basic features of tight-binding lattice models for Weyl semimetals are reviewed and the transfer matrix is constructed analytically. In \S \ref{sec:spcl}, a particular family of 4-node models is discussed with tunable connectivity of the Fermi arcs. In \S \ref{sec:emresponse}, we compute the polarization/magnetization response for our 4-node WSM model Hamiltonians. We close with some comments and implications in \S  \ref{sec:concl}. The details of the transfer matrix calculations have been relegated to Appendix \ref{app:tmat}. 

A comment on notation: We use the upright fonts ($\mathrm{T}$, $\mathrm{I}$, etc) to denote the action of symmetry operation on our system, including the action on spacetime, and the corresponding calligraphic font ($\mathcal{T}$, $\mathcal{I}$, etc) to denote its action on the internal (pseudospin) degrees of freedom.


\section{WSM Generalities}    \label{sec:wsm}

\subsection{Lattice models}
We consider a general class of 2-band lattice models described by the Bloch Hamiltonian
\beq 
  \hlt(\vk) = \hlt_x(k_x) + \hlt_\perp(\vk_\perp),   \label{eq:hlt_orig} 
\eeq 
where 
\begin{align} \label{eq:4nodewsm1}
  \hlt_x(k_x) = & \; \sin k_x \sigma^x + (1 - \cos k_x) \sigma^z, \nonumber \\ 
  \hlt_\perp(\vk_\perp) = & \; \gamma(\vk_\perp) \id + \hlts_y(\vk_\perp) \sigma^y + \hlts_z(\vk_\perp) \sigma^z, 
\end{align}
$\gamma, \hlts_y, \hlts_z : \torus^2 \to \real$ are functions of the transverse (``surface'') momentum $\vk_\perp \equiv (k_y, k_z) \in \torus^2$. The Pauli matrices correspond to orbital/sublattice degrees of freedom (hereafter \emph{pseudospin}), so that the time reversal operator satisfies $\trev^2 = \id$. 

Along the $x$ direction, the model is described by a lattice version of the 1D Dirac Hamiltonian, with $1 - \cos k_x$ being the UV regulator. In condensed matter terminology, it is a model of a 1D topological insulator tuned to the gapless point, so that one may obtain edge states when $\hlts_z(\vk_\perp) < 0$ and $\hlts_y(\vk_\perp) = 0$. 
 
In Table \ref{tb:symm}, we list the discrete symmetry operators for the Hamiltonian, as well as the conditions on $\gamma, \hlts_y, \hlts_z$ for the Hamiltonian to be invariant under them. We note that retaining both $\trev$ and $\inv$ demands that $\hlts_y = 0$.

\begin{table}[htb!]
  \vspace{0.1in}
  \renewcommand{\arraystretch}{1.5}
  \setlength{\tabcolsep}{5pt}
  \newcolumntype{L}{>{\centering\arraybackslash}m{0.2\columnwidth}}
  \begin{tabular}{c|LLL}
    \hline 
    Symmetry  & Time reversal & Inversion & Charge conjugation\\ 
    \hline
    Operator & $\trev = \sigma_z K$ & $\inv = \sigma_z$ & $\cc = \sigma_x K$ \\
    \hline
    $\gamma(\vk_\perp)$ & even & even & odd\\
    $\hlts_y(\vk_\perp)$ & even & odd & odd \\
    $\hlts_z(\vk_\perp)$ & even & even & even \\ 
    \hline  
  \end{tabular}
  \caption{Symmetries of the Hamiltonian in \eq{eq:hlt_orig}. The Hamiltonian is symmetric under a given symmetry if the corresponding element $\gamma, \hlts_y, \hlts_z$ are odd/even under $\vk_\perp \to -\vk_\perp$.}   \label{tb:symm}
\end{table}

The spectrum of $\hlt(\vk)$ is given by 
\beq \ve(\vk) = \gamma \pm \sqrt{\sin^2 k_x + \hlts_y^2 + (1 - \cos k_x + \hlts_z)^2}. \eeq
Assuming $\hlts_z(\vk_\perp) > -2$, as we shall do throughout this paper, the bulk gap closes for $k_x = 0$ at $\vk_\perp$ satisfying
\beq 
\hlts_y(\vk_\perp) = \hlts_z(\vk_\perp) =  0.   \label{eq:nodes}
\eeq 
Near a gapless point $\vk = \vk^\ast$, the Hamiltonian becomes 
\beq  
  \hlt(\vk^\ast + \delta \vk) = \delta k_i V_{ij}(\vk^\ast) \sigma^j + O(\delta k^2), 
\eeq 
which corresponds to a Weyl node if $\det V \neq 0$, with its chirality given by $\chi(\vk^\ast) = \sgn{\det V(\vk^\ast)}$. Explicitly, for $\hlt(\vk)$ defined in \eq{eq:hlt_orig}, 
\beq 
\det V(\vk^\ast) = \left. \frac{\partial(\hlts_y, \hlts_z)}{\partial(k_y, k_z)} \right|_{\vk = \vk^\ast}, \label{eq:chirality}  
\eeq 
i.e, the Jacobian of $(\hlts_y, \hlts_z)$. We also note that the Weyl nodes occurs at energy $\ve = \gamma(\vk^\ast)$, so that they can be at different energies for $\gamma \neq 0$.

If we have Weyl nodes in the $k_x = 0$ plane, then there will be nontrivial surface modes for a surface normal to $x$. Thus, we should to study our system on a slab geometry, finite along the $x$-axis and infinite (or periodic) along the $y$- and $z$-axes, so that $\vk_\perp$ is a still good quantum number. We seek the values of $\vk_\perp$ for which there is a localized surface mode, as well as the corresponding energies. We next compute these using transfer matrices.

\subsection{Transfer matrices} 
\label{subsec:tm}
Transfer matrices naturally arise in the study of finite order linear difference equations, an instance of which is the Schr\"odinger equation for 1D tight binding models with finite range hoppings. They are operators that translate the wavefunctions by a finite distance. Transfer matrices are useful in studying finite 1D chains as an analysis of their spectra reveals the delocalized (``bulk'') modes as well as the modes localized on the edge. 

To construct a transfer matrix for translations along $x \in  [0, L]$, we inverse Fourier transform the Hamiltonian of \eq{eq:hlt_orig} along $x$ to write it as a set of one-dimensional chains, one for each value of $\vk_\perp$. Explicitly, 
\begin{align}
  \hlt(\vk_\perp) = & \; \sum_{n}  \bigg[  \vcd_{n+1} \left( \frac{ \sigma^x - i \sigma^z}{2} \right) \vc_n + \text{h.c} \nonumber \\ 
  & + \vcd_n \left( \gamma \id + \hlts_y \sigma^y + (1 + \hlts_z) \sigma^z \right) \vc_n \bigg],  
\end{align}
where $\vcd$, $\vc$ are 2-component fermionic creation/annihilation operators corresponding to the pseudospin. The Schr\"odinger equation, $\hlt \ket{\Psi} = \ve \ket{\Psi}$, can then be written as a recursion relation
\beq J \Psi_{n+1} + M \Psi_n + J^\dagger \Psi_{n-1} = \ve \Psi_n, \eeq 
where $\Psi_n(\vk_\perp) = \bra{\Omega} \vc_n \ket{\Psi(\vk_\perp)}$ denotes the 2-component wavefunction at site $n$ along $x$ for a given $\vk_\perp$, and $\ket{\Omega}$ denotes the fermionic vacuum. The \emph{hopping} ($J$) and \emph{on-site} ($M$) matrices can be identified as
\begin{align}
  J = & \; \frac{1}{2i} \left( \sigma^x - i \sigma^z \right), \nonumber \\ 
  M = & \; \gamma \id + \hlts_y \sigma^y + (1 + \hlts_z) \sigma^z.
\end{align}

Clearly, $J$ is singular. Using the methods proposed in Ref. \onlinecite{vd-vc_tm} to construct a transfer matrix for systems with a noninvertible hopping matrix, we derive (for details, see Appendix \ref{app:tmat}):
\beq 
T(\ve, \vk_\perp) = \frac{1}{1 + \hlts_z}\left( \begin{array}{cc} (\ve - \gamma)^2 - \Lambda^2 & \quad  -(\ve - \gamma - \hlts_y) \\ \ve  - \gamma + \hlts_y & \quad -1 \end{array} \right), 
\eeq
where $\Lambda^2 = \hlts_y^2 + (1 + \hlts_z)^2$. 

The system exhibits surface states localized at $x = 0, L$, hereafter termed the ``left'' and ``right'' edge states, respectively. For the left edge, we use the Dirichlet boundary condition and demand that $\Phi_1 = (1,0)^T$ be an eigenvector of the transfer matrix with eigenvalue inside the unit circle (Appendix \ref{app:tmat}). Thus, for the left surface, we get a localized mode with energy:
\beq 
  \ve_L (\vk_\perp) = \gamma(\vk_\perp) - \hlts_y(\vk_\perp), \quad \text{if } \, \hlts_z(\vk_\perp) < 0.  \label{eq:spec_edgeL_orig}
\eeq 
Similarly, for the right surface, we get 
\beq 
  \ve_R (\vk_\perp) = \gamma(\vk_\perp) + \hlts_y(\vk_\perp), \quad \text{if } \, \hlts_z (\vk_\perp) < 0.  \label{eq:spec_edgeR_orig}
\eeq

For $\gamma(\vk_\perp) = 0$, i.e, the Weyl nodes being at the same energy, the Fermi arcs are defined as the set of surface momenta $\vk_\perp$ for which there is a localized state with $\ve_{edge}(\vk_\perp) = 0$.  Hence, on the left surface, the Fermi arcs are simply the loci of $\vk_\perp \in \torus^2$ satisfying
\beq 
\hlts_y(\vk_\perp) = 0, \quad   \hlts_z(\vk_\perp) < 0.    \label{eq:spec_edgeL}
\eeq
while on the right surface, they are the loci of
\beq 
\hlts_y(\vk_\perp) = 0, \quad   \hlts_z(\vk_\perp) < 0.    \label{eq:spec_edgeR}
\eeq 
Clearly, the Fermi arcs end at the projections of the Weyl nodes on the surface, given by $\hlts_y(\vk_\perp) = 0 = \hlts_z(\vk_\perp)$.

Generically, $\eta_y = 0$ and $\eta_z = 0$ describe 1-dimensional curves in the surface Brillouin zone $\torus^2$, with the Weyl nodes lying at their intersection and the Fermi arcs lying along the former. In the next section, we engineer a one parameter family of curves whose points of intersection are independent of the parameter.


\section{Specific lattice models}    \label{sec:spcl}

\subsection{2 node WSM}
\begin{figure*}[htb]
\centering
  \begin{tabular}{cc}
    \parbox[c][][c]{0.2\textwidth}{
      \ \\ \vspace{-0.02in} 
      \includegraphics[width=0.20\textwidth]{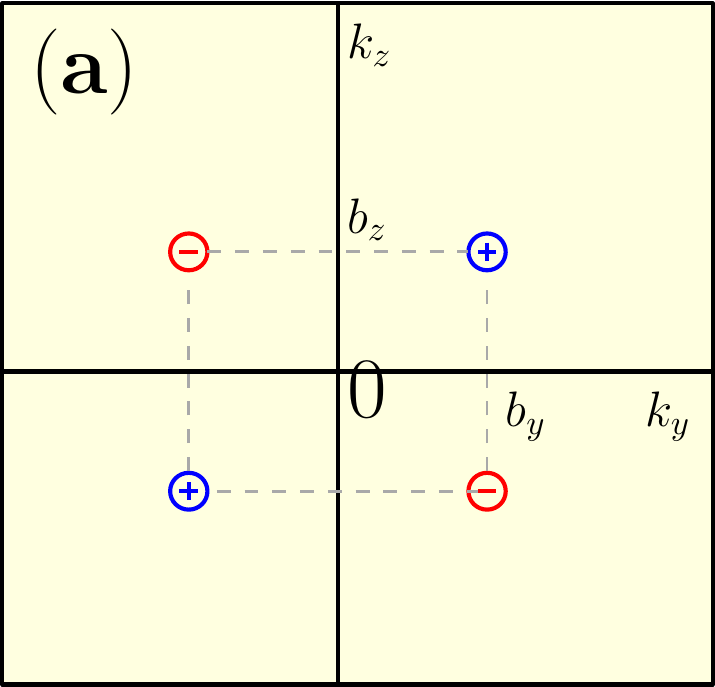} \\ \vspace{0.04in}
      \includegraphics[width=0.20\textwidth]{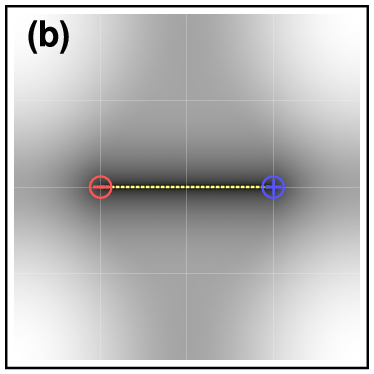}  
    }  
    & 
    \parbox[c][][c]{0.8\textwidth}{
      \includegraphics[width=0.8\textwidth]{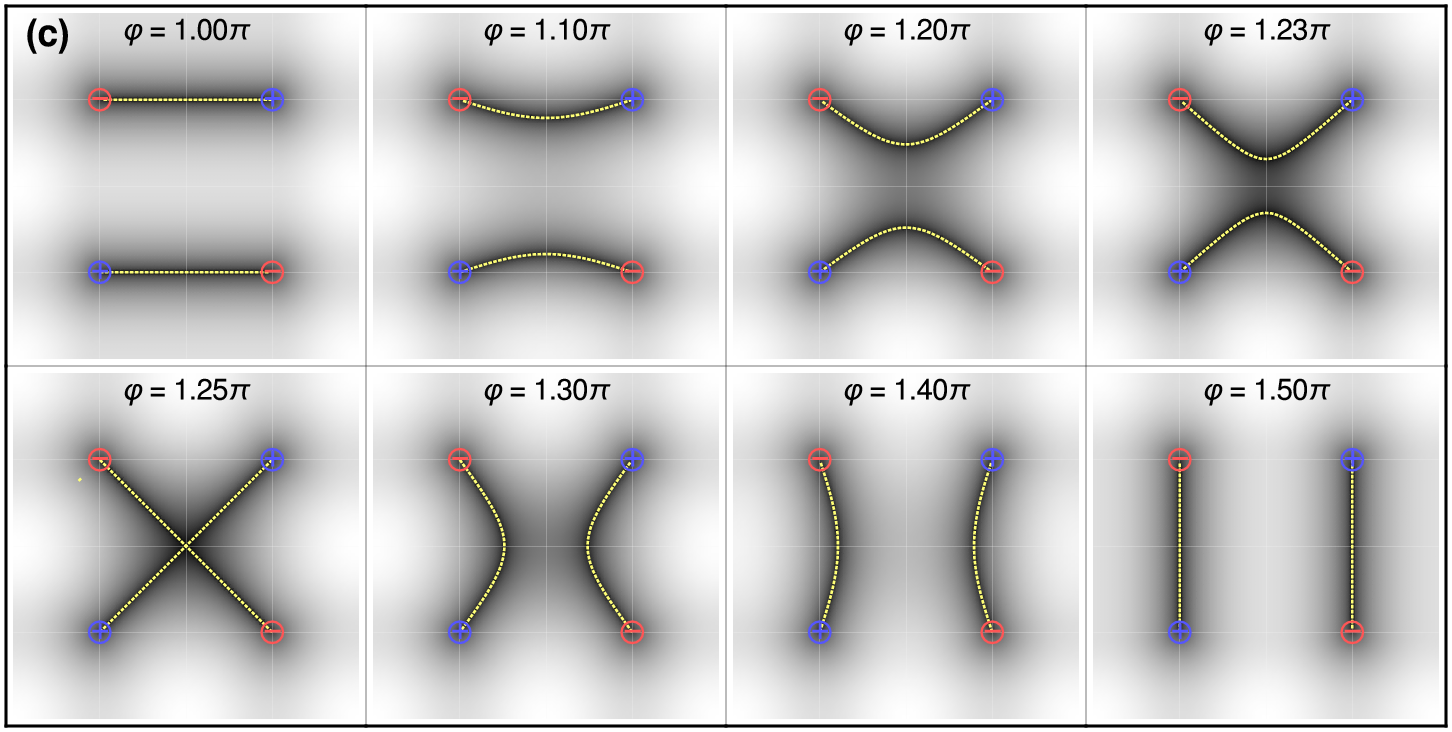} 
    }
    
  \end{tabular}
  \caption{  
    (color online) (a) The schematic of Weyl nodes for the 4-node model of \eq{eq:4nodewsm2}. The analytically computed Fermi arcs (yellow, dashed line), overlaid on the surface spectrum for of the surfaces computed using ED, for (b)the 2-node model of \eq{eq:2node_wsm}, with nodes at $\vk_\perp^\ast = \left( \pm \frac{\pi}{2},0 \right)$, and (c) the 4-node model Hamiltonian  of \eq{eq:4nodewsm3} with the nodes at $\vk_\perp^\ast = \left( \pm \frac{\pi}{2}, \pm \frac{\pi}{2} \right)$, as we tune $\varphi$. The analytic and numerical computations are in good agreement, as expected. 
  }
  \label{fig:fermi_arcs} 
  \vspace{-0.15in}
\end{figure*}

We start off by demonstrating the transfer matrix calculations for a simple 2-band, 2-node model Hamiltonian for a Weyl semimetal\cite{hughes-srinidhi_emres}, given by
\begin{align}
 \hlt = & \; \sin k_x \sigma^x + \sin k_y \sigma^y \nonumber \\ 
 & \; + (2 + \cos b_z - \cos k_x - \cos k_y - \cos k_z) \sigma^z.   \label{eq:2node_wsm}
\end{align}
We identify
\begin{align}
 \hlts_y (k_y, k_z) = & \; \sin k_y, \nonumber \\ 
 \hlts_z (k_y, k_z) = & \; 1 + \cos b_z - \cos k_y - \cos k_z,
\end{align}
As $\hlts_y$ is odd and $\hlts_z$ even under $\vk_\perp \to -\vk_\perp$, the model is symmetric under inversion and charge conjugation (see Table \ref{tb:symm}). The Weyl nodes are given by 
\beq k_x = k_y = 0, \quad k_z = \pm b_z. \eeq 
The Fermi arcs stretch between these nodes along 
\beq 0 = \hlts_y(\vk_\perp) = \sin k_y \implies k_y = 0, \eeq 
for 
\beq -2 < \cos b_z - \cos k_z < 0 \implies k_z < b_z. \eeq 
A plot of this calculation of the Fermi arc, superposed on the surface spectrum obtained from exact diagonalization, is shown in Fig. \ref{fig:fermi_arcs}(b).

\subsection{4 node WSM}
In this section, we describe a family of lattice models defined as
\beq  
\label{eq:4nodewsm2}
\vecenv{\hlts_y}{\hlts_z} = M \vecenv{\cos k_y - \cos b_y}{\cos k_z - \cos b_z},   
\eeq
where $M \in \SL(2, \real)$.  Using \eq{eq:nodes} and the fact that $M$ is invertible, we get four Weyl nodes at $k_y = \pm b_y, k_z = \pm b_z$, with chiralities
\begin{align}
 \chi = & \; \det M(\varphi) \;  \sin k_y \sin k_z \big|_{\vk = \vk^\ast} \nonumber \\ 
 = & \; \sin k_y^\ast \sin k_z^\ast. 
\end{align}
Clearly, the location and chiralities of the Weyl nodes is independent of the choice of $M$ (see Fig \ref{fig:fermi_arcs}(a)).

As $\hlts_y$ and $\hlts_z$ are both even under $\vk_\perp \to -\vk_\perp$, the model is symmetric under time-reversal (see Table \ref{tb:symm}), as well as under reflections about $k_y$ and $k_z$ axes. Since time reversal symmetry demands that given a Weyl node at $\vk = \vk^\ast$, there must be another one of the same chirality at $\vk = -\vk^\ast$, the Weyl nodes must lie at the vertices of a parallelogram centered at zero. 

On the other hand, the Fermi arcs, given by $\hlts_y = 0$, depend strongly on the choice of $M$. To study them more explicitly, we set $b_y = b_z = \pi/2$ and consider only the matrices $M(\varphi) \in SO(2) \subset SL(2, \real)$, so that explicitly
\beq 
\label{eq:4nodewsm3}
\vecenv{\hlts_y}{\hlts_z} = \left( 
\begin{array}{cc} \cos\varphi & \;\; -\sin\varphi \\ \sin\varphi & \;\; \;\;\; \cos\varphi \end{array}  
\right) \vecenv{\cos k_y}{\cos k_z}. 
\eeq 
Thus, $M(\varphi), \; \varphi \in [0, 2\pi)$ implements a clockwise rotation by $\varphi$ on $\real^2$. By definition, $\hlts_z > -2$, since 
\begin{align*}
 \hlts_z = & \; \sin\varphi \cos k_y + \cos \varphi \cos k_z \\
 = & \sqrt{\cos^2 k_y + \cos^2 k_z}  \cos\left( \varphi - \tan^{-1} \left( \frac{\cos k_y}{\cos k_z} \right) \right) \\
 \geq & - \sqrt{\cos^2 k_y + \cos^2 k_z}  \geq - \sqrt{2}. 
\end{align*}
For $b_y, b_z$ sufficiently far from $\pi/2$, this condition may be violated, so that one might end up nucleating extra Weyl nodes in the $k_x = \pi$ plane as one tunes  $\varphi$. 

The Fermi arcs are explicitly given by the equation
\beq 
  \hlts_y = \cos\varphi \cos k_y - \sin \varphi \cos k_z = 0. 
\eeq 
We plot some of these Fermi arcs as a function of $\varphi$, compared with the exact diagonalization results, in Fig.~\ref{fig:fermi_arcs}(c). Analytically, we note that the Fermi arcs run along the $k_y$ axis for $\varphi = \left( n + \frac{1}{2} \right) \pi$ and along the $k_z$ axis for $\varphi = n \pi$. Another special case of interest is $\varphi = \left( n + \frac{1}{4} \right) \pi$, when the Fermi arcs run along $k_y = \pm k_z$ and hence we have degenerate zero energy modes at surface momenta $\vk_\perp = (0,0)$ or $(\pi,\pi)$. 

\begin{figure}[tb!]
 \includegraphics[width=\columnwidth]{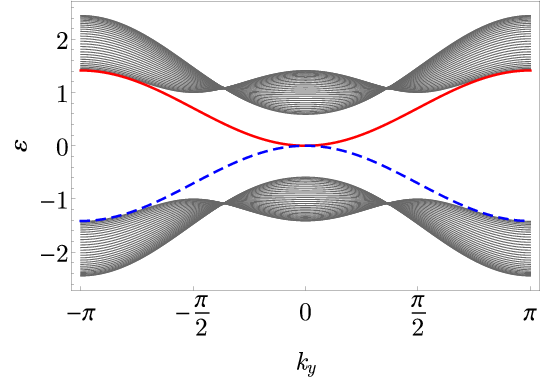}  
 \caption{
    (color online) The spectrum  of the Hamiltonian of \eq{eq:4nodewsm3} on a slab, finite along the $x$-direction with $k_z = 0$, as a function of $k_y$. The red(solid bold) and blue(dashed) lines highlight the spectra for the left and right surfaces, respectively, which clearly touch $\ve = 0$ quadratically.
  }
 \label{fig:ED}
  \vspace{-0.15in}
\end{figure}

Generically, we do not expect such a degeneracy to be stable, as adding a surface ``mass'' term would lead to the Fermi arcs splitting into an avoided crossing. However, the degeneracy can be protected by a lattice symmetry, specifically, the 4-fold rotoreflection symmetry in the $y$-$z$ plane, corresponding to the crystallographic point group symmetry group  $\mathrm{S}_{4}$. Note that this is not same as the group of permutations, confusingly also denoted by $\mathrm{S}_4$. Explicitly, the $\mathrm{S}_4$ symmetry acts as
\beq
\hlt(k_x, \vk_\perp) \longmapsto  \mathcal{S}_4 \cdot \hlt(-k_x, \mathcal{R}_{\pi/2} \cdot \vk_\perp) \cdot \mathcal{S}^{-1}_4,    \label{eq:C4v_def}
\eeq 
where $\mathcal{S}_4 = \sigma^z$ and $\mathcal{R}_{\theta}$ is the orthogonal matrix representing a $\theta$ rotation in $\real^2$. For $\varphi = \pi/4, 5\pi/4$ (Fig. \ref{fig:fermi_arcs}(e)), the bulk Hamiltonian becomes 
\begin{align}
  \hlt(\vk) = & \; \sin k_x \sigma^x \pm \frac{1}{\sqrt{2}} \left( \cos k_y - \cos k_z \right) \sigma^y \nonumber \\ 
  + & \left( (1-\cos k_x) \pm \frac{1}{\sqrt{2}} \left( \cos k_y + \cos k_z \right) \right) \sigma^z,
\end{align}
which is clearly invariant under \eq{eq:C4v_def}. Expanding the edge spectrum near the crossing point $\vk_\perp = 0$ for  $\varphi = 5\pi/4$, we get
\begin{align} 
\ve_{edge}(\vk_\perp) = & \; -\frac{1}{\sqrt{2}} \left( \cos k_y - \cos k_z \right) \nonumber \\ 
= & \; \frac{1}{2\sqrt{2}} \left( k_y^2 - k_z^2 \right) + O(k_\perp^4).
\end{align}
Thus, the surface spectrum touches $\ve = 0$ quadratically at the $\mathrm{C}_4$-symmetric point $\vk_\perp = 0$. This is analogous to the case of topological crystalline insulators\cite{fu_TCI}, where one gets a quadratic band-touchings protected by a lattice symmetry at the points in the Brillouin zone symmetric under that lattice symmetry. In Fig \ref{fig:ED}, we plot the edge spectrum for $\varphi = 5\pi/4$ along $k_z = 0$ computed using exact diagonalization, which clearly exhibits the ``quadratic band touching'' behavior.

In conclusion, we have constructed a family of models which are identical in terms of their location of the Weyl nodes and hence the low energy behavior, but display dramatic differences in their Fermi arcs. Furthermore, the basic construction of \eq{eq:4nodewsm2} can be readily generalized to a bigger class of Weyl semimetals with tunable arcs, by setting
\beq 
  \vecenv{\hlts_y}{\hlts_z} = M(\varphi) \vecenv{f_y}{f_z},
\eeq 
where $f_y(\vk_\perp)$, $f_z(\vk_\perp)$ are arbitrary functions of $\vk_\perp$, even under $\vk_\perp \to -\vk_\perp$.  The nodes are then independent of $\varphi$, varying which one can tune the Fermi arcs continuously.

\section{Electromagnetic response}    \label{sec:emresponse}
The most common feature of WSMs' universal transport characteristics is the anomalous Hall effect(AHE), with Hall coefficient
\beq 
  \sigma_H = \frac{\ve^2}{2\pi h} \sum_i \chi_i \vk^\ast_i, 
\eeq 
which can be derived from an effective action at the linear order in $\vk - \vk^\ast$. However, the anomalous Hall coefficient clearly vanishes for our model, so that there is no ``quasi-topological'' contribution to the transport. This also follows from the fact that our model is symmetric under time reversal. 

Furthermore, we note that at the linear order, the effective action depends only on the positions and chiralities of the Weyl nodes, which are independent of the parameter $\varphi$ in our model. Thus, any ``topological'' contribution to the linear response derived from the effective action must be independent of $\varphi$. However, the edge states, which can give us a nonzero polarization/magnetization, depend strongly on $\varphi$. In the following, we show that the response indeed depends on $\varphi$.

\paragraph{Polarization:} 
We compute the polarization along $x$ using two independent methods: 
\begin{enumerate}
 \item Using ED, we compute the excess in the number of occupied edge states between opposite edges. 
 \item Using the exact surface spectra obtained from the transfer matrix calculations, we compute the charge accumulated at each surface. 
\end{enumerate}
For the latter, we note that each edge state for a given $\mathbf{k}_{\perp}$ can be thought of as the end of a 1D wire, and hence contributes a total of $\frac{e}{2}$ to the charge polarization (assuming normal ordering). In the ground state at $T = 0$, the Fermi-Dirac statistics is given by 
\[ f_{FD} (\ve, T = 0) = \Theta(\mu - \ve). \]
Thus, all states with $\ve_{L,R} < \mu$ are filled up, where we set $\mu = 0$, the energy of the Weyl nodes. Using \eq{eq:spec_edgeL} and \eq{eq:spec_edgeR}, the $\vk_\perp$ for which we have a filled state are given by $\hlts_y(\vk_\perp) > 0, \hlts_z(\vk_\perp) < 0$ for the left surface and by $\hlts_y(\vk_\perp) < 0, \hlts_z(\vk_\perp) < 0$ for the right surface. We can compute the polarization as 
\begin{align}
  P_{x}(\varphi) = \frac{e}{2} \int_{\torus^2} \frac{d\mathbf{k}_{\perp}}{(2\pi)^{2}} \Theta(-\eta_{z})\left\lbrack \Theta(\eta_{y})-\Theta(-\eta_{y})\right\rbrack.
\label{eq:areapol}
\end{align}

\begin{figure*}[htb]
\centering  
  \includegraphics[width=1.7\columnwidth]{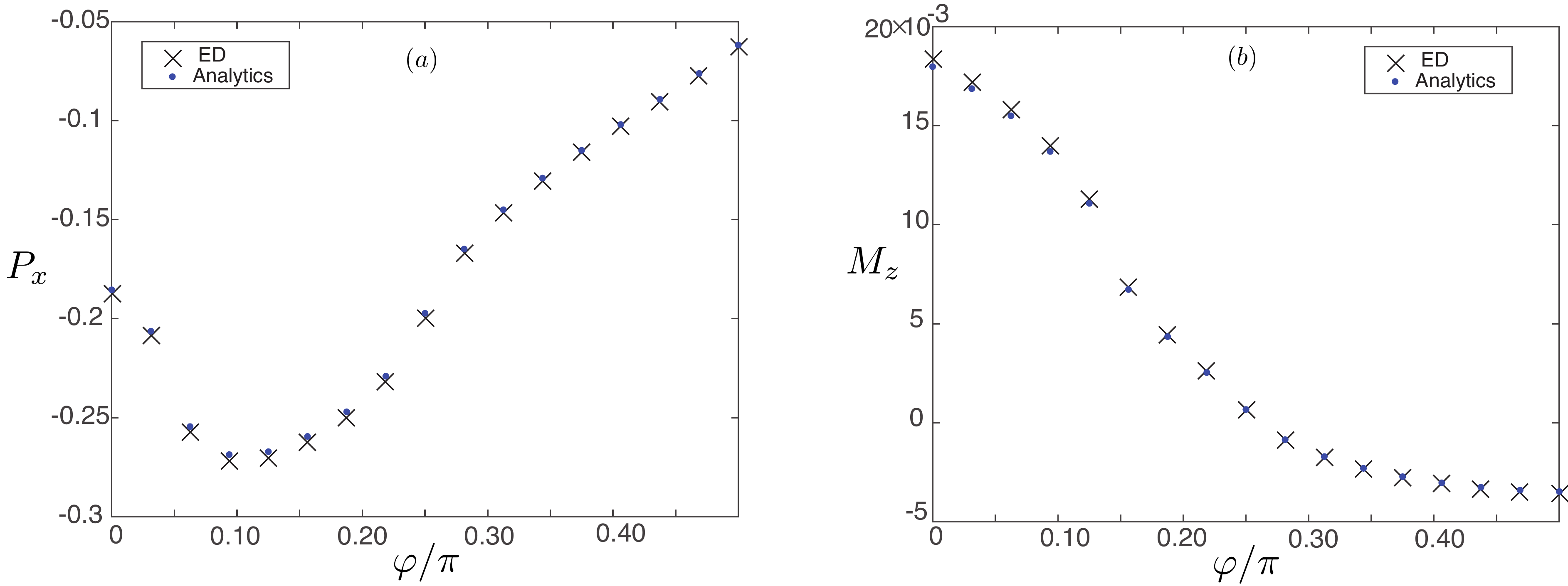} 
  \caption{
    The $\varphi$-dependent EM response for the model Hamiltonian defined in eqs.~\ref{eq:4nodewsm1},\ref{eq:4nodewsm2}, with $b_{y}=b_{z}=0.38\pi$, both computed in two different ways (\S \ref{sec:emresponse}). (a) The polarization in the $x$-direction, $P_{x}$, calculated from ED as well as computing the surface charge(\eq{eq:areapol}) numerically (labelled `Analytics' in the plot). (b) The magnetization along $z$, $M_z$, defined in \eq{eq:areamag}, with the currents calculated from ED as well as using the analytic form of the surface spectrum(\eq{eq:spec_edgeL_orig}, \ref{eq:spec_edgeR_orig}). All ED computations were performed on a $100\times 100\times 100$ grid. There is an excellent agreement between the two calculations for both (a) and (b).
  }
  \label{fig:comparison}
\end{figure*}

\paragraph{Magnetization}
In order to get a nonzero magnetization, there must be Weyl nodes of opposite chiralities at different energies, which can be achieved by adding a nonzero $\gamma(\vk_\perp)$. Turning on this $\gamma(\vk_\perp)$ adiabatically, we would expect boundary currents to arise. We again compute the current using two independent methods:
\begin{enumerate}
 \item Using ED, we compute the total current in the ground state of the system. 
 \item Using the exact surface spectra obtained from the transfer matrix calculations, we compute the total surface current. 
\end{enumerate}
To derive the surface current from the surface spectrum, consider a system with Hamiltonian $\mathcal{H}(\mathbf{k})$ minimally coupled to an external electromagnetic field as $\mathbf{k}\rightarrow \mathbf{k}+e\mathbf{A}$. The current operator is defined as:
\begin{align}
\mathcal{J}^{a} \equiv & \; \frac{\delta \mathcal{H}(\mathbf{k}+e\mathbf{A})}{\delta A_{a}}\Bigg\vert_{\mathbf{A}\rightarrow 0} \\ 
= & \; e \sum_{k_{a}, \alpha, \beta} \frac{\partial \mathcal{H}^{\alpha\beta}(\mathbf{k})}{\partial k_{a}} c^{\dagger}_{k_{a},\alpha} c^{\phantom{\dagger}}_{k_{a},\beta}.
\end{align} 
We can now take the expectation value of the current operator over the ground state many-body wavefunction to get the current in the ground states, $J^a \equiv \langle \mathcal{J}^a \rangle$. We compute this by taking a trace over the occupied states:
\begin{align} 
  J^{a} = \frac{e}{2} \sum_{n} \int_{\torus^2} \frac{d\vk_\perp}{(2\pi)^2} \frac{\partial \ve_{n}(\vk_\perp)}{\partial k_{a}} f_{FD}(\ve(\vk_\perp), T)
\end{align} 
At $T = 0$, the total ground state current is obtained by simply summing over all occupied state, i.e, all states with $\ve < \mu = 0$.  Explicitly, the edge state dispersion on the surfaces normal to $\hat{x}$ is given by 
\[ 
  \ve_{edge}(\vk_\perp) = \gamma(\mathbf{k}_{\perp})\pm \hlts_y(\vk_\perp). 
\] 
Since $\hlts(\vk_\perp)$ is even under $\vk_\perp \to -\vk_\perp$, its derivative must be odd, which would integrate out to zero, as the rest of the integrand is even. This is also expected on physical grounds, as the current should vanish in the ground state when $\gamma = 0$. 

Thus, on the left surface, the current along the $a=y,z$ directions when $\gamma(\mathbf{k}_{\perp})$ is turned on adiabatically is
\begin{align}
\label{eq:areamag}
  J^{a}=\frac{e}{2} \int_{\torus^2}  \frac{d\mathbf{k}_{\perp}}{(2\pi)^{2}} \frac{\partial \gamma(\mathbf{k}_{\perp})}{\partial k_{a}} \Theta(-\eta_{z}) \left\lbrack\Theta(\eta_{y})-\Theta(-\eta_{y})\right\rbrack.
\end{align} 
The right surface has an overall minus sign in the current since the pre-factor due to the charge would be $-\frac{e}{2}$. Finally, a surface current $J_{y}$ on the surface perpendicular to $\hat{x}$ gives rise to a magnetization $M_{z}$ and vice versa. 

In Fig.~\ref{fig:comparison}, we plot the polarization and magnetization as a function of $\varphi$, computed from ED as well as using eq.~\ref{eq:areapol} for $b_{y} = b_{z} = 0.38\pi$. The EM response obtained from ED are in excellent agreement with those obtained using the analytic expression for the surface spectrum.

\section{Discussion and Conclusions}   \label{sec:concl}
The conventional paradigm for calculating the EM response of a ``topological'' semimetal involves writing down a linearized (low energy) model for the bulk and calculating an effective action for the electromagnetic field by means of dimensional reduction or a direct Feynman diagrammatic calculation. It is \emph{believed} that such an effective action captures all the ``universal'' features of the EM response, i.e, the features that are not affected by an addition of boundary terms. However, in this paper, we construct an explicit counterexample to this \emph{belief}, where the universal transport properties are not completely characterized by this simplistic approach.

Keeping the nodes fixed in the bulk, i.e. starting with the same linearized model, we have engineered different Fermi arc configurations that are not completely characterized by the bulk low energy physics. This has a direct measurable consequence in the EM response such as polarization/magnetization of the system. Thus, the microscopics of the system, i.e. knowledge of the lattice model is \emph{indispensable} in predicting these properties. 

Our statement about the charge polarization/magnetization \emph{for a semimetal} merits some explanation. Na\"ively, one might think that a \mbox{(semi-)metal} cannot sustain a nonzero charge polarization, as any accumulation of charge on a surface can be neutralized by a current flowing through the bulk. However, for Weyl semimetals, the bulk single particle states connecting the opposite surfaces exist only for a finite set of lattice momenta. Thus, for a translation invariant system, one may have localized charges at a given surface, as one cannot scatter to the modes extended in the bulk. 

Secondly, we have resorted to a real space calculation to compute the polarization and the magnetization. For gapped systems, these calculations should give the same result as a momentum space calculation involving an integral over the Berry connection. However, for gapless systems,  integrals over the entire Brillouin zone involving the Berry connection are potentially divergent. This is because the Berry connection has a singularity at the gapless points. Furthermore, a WSM has a line of poles (Dirac strings) running between pairs of Weyl nodes of opposite chirality, which are dependent on a gauge choice for the Berry connections. Thus, in order to compute the divergent polarization and magnetization integrals over the momentum space, one must choose a suitable regularization for the integrals to obtain a finite result. We leave the choice of a regulator, potentially based on a physical principle to match with the real space calculations, for a future work. 

We highlight an interesting feature of our model by thinking of the Berry monopoles as isolated ``\emph{magnetic charges}'', which must add up to zero (fermion doubling). Most of the models for Weyl semimetals studied so far in the literature have a nonzero \emph{dipole moment} for this \emph{charge} configuration, and the AHE coefficient is proportional to this \emph{dipole moment}. However, we present a model where these \emph{charges} form a pure quadrupole, for which the AHE coefficient vanishes. However, it would exhibit the nonuniversal nonlinear response corresponding to a Berry quadrupole, as described by Liang Fu in Ref.~\onlinecite{fu_quadrupole}\footnote{Somewhat strangely, he refers to the configuration as a \emph{Berry dipole} as opposed to a \emph{Berry quadrupole}.}. An interesting extension of this picture would be to compute the response for a generic nodal semimetal in a ``multipole expansion'', analogous to the conventional electrostatic case. 

The study of geometrical and topological aspects of conventional band theory has led to many interesting ideas and discoveries in condensed matter physics in the recent decades. A particularly profound example is the idea of bulk-boundary correspondence for gapped phases, which has not been satisfactorily studied for gapless phases. We hope that this work would further the understanding of the bulk-boundary connection for WSMs.

\acknowledgements
We acknowledge useful conversations with Victor Chua, Awadhesh Narayan, Apoorv Tiwari, Ashvin Vishwanath, Taylor Hughes and Adolfo Grushin. VD was supported by the National Science Foundation through Grant NSF DMR 13-06011. STR was supported by the Office of Naval Research through Grant ONR YIP Award N00014-15-1-2383.


\appendix 
\section{Calculating the transfer matrix}    \label{app:tmat}
We follow the method of Ref.\onlinecite{vd-vc_tm} to compute the transfer matrix for our general model for a WSM. For the recursion relation
\beq 
J \Psi_{n+1} + M \Psi_n + J^\dagger \Psi_{n-1} = \ve \Psi_n,  \label{eq:recur}
\eeq 
we identified 
\begin{align}
  J = & \; \frac{1}{2i} \left( \sigma^x - i \sigma^z \right), \nonumber \\ 
  M = & \; \gamma \id + \hlts_y \sigma^y + (1 + \hlts_z) \sigma^z. 
\end{align}
Clearly, $\rank(J) = 1$, $J^2 = 0$ and the reduced singular value decomposition of $J$ is $J = \vv \cdot \vw^\dagger$, with
\beq 
\vv = \frac{1}{\sqrt{2}} \vecenv{-i}{1}, \quad  \vw =  \frac{1}{\sqrt{2}} \vecenv{i}{1}, 
\eeq 
which satisfy
\[ \vv^\dagger \vv = \vw^\dagger \vw = 1, \quad \vv^\dagger \vw = 0. \] 
The on-site Green's function is 
\begin{align}
  \green = & \;  (\ve \id - M)^{-1} \nonumber \\ 
  = & \; \frac{1}{(\ve - \gamma)^2 - \Lambda^2} \left[ (\ve - \gamma) \id + \hlts_y \sigma^y + (1 + \hlts_z) \sigma^z  \right]
\end{align}
where $\Lambda^2 = \hlts_y^2 + (1 + \hlts_z)^2$. Its restrictions to the $\vv,\vw$ subspaces are  
\begin{align}
 \green_{vv} = & \; \vv^\dagger \green \vv = \frac{\ve - \gamma + \hlts_y}{(\ve - \gamma)^2 - \Lambda^2}, \nonumber \\ 
 \green_{ww} = & \; \vw^\dagger \green \vw = \frac{\ve - \gamma - \hlts_y}{(\ve - \gamma)^2 - \Lambda^2}, \nonumber \\
 \green_{vw} = \green_{wv}^\ast = & \; \vw^\dagger \green \vv = -\frac{1 + \hlts_z}{(\ve - \gamma)^2 - \Lambda^2}.
\end{align}

The transfer matrix construction follows from expressing $\Psi_n = \alpha_n \vv + \beta_n \vw, \alpha_n, \beta_n \in \cmplx$, and extracting the coefficients $\alpha_n$ and $\beta_n$ in \eq{eq:recur} as 
\begin{align}
 \alpha_n = & \; \green_{vv} \beta_{n+1} + \green_{wv} \alpha_{n-1}, \nonumber \\
 \beta_n = & \; \green_{vw} \beta_{n+1} + \green_{ww} \alpha_{n-1}, \nonumber 
\end{align}
which can be rearranged to give 
\beq \Phi_{n+1} = T \Phi_n, \quad \Phi_n = \vecenv{\beta_n}{\alpha_{n-1}}, \eeq
where
\begin{align}
  T = & \; \frac{1}{|\green_{vw}|} \left( \begin{array}{cc} 1 & \quad  -\green_{ww} \\ \green_{vv} & \quad |\green_{vw}|^2 - \green_{vv} \green_{ww} \end{array} \right) \nonumber \\ 
  = & \; \frac{1}{1 + \hlts_z}\left( \begin{array}{cc} (\ve - \gamma)^2 - \Lambda^2 & \quad  -(\ve - \gamma - \hlts_y) \\ \ve  - \gamma + \hlts_y & \quad -1 \end{array} \right). 
\end{align}
One can explicitly check that $\det T = 1$, so that its eigenvalues are
\[
\rho = \frac{1}{2} \left( - \tr T \pm \sqrt{\left( \tr T \right)^2  - 4}  \right). 
\] 
Clearly, $\rho \in \real$ for $|\tr T| > 2$, corresponding to growing/decaying states, and $\rho$ lies on the unit circle for $|\tr T| < 2$, which corresponds to Bloch states. The band edges are given by $|\tr T| = 2$, i.e, 
\[  
\left| \frac{(\ve-\gamma)^2 - \Lambda^2 - 1}{1 + \hlts_z} \right| = 2,
\] 
which can be solved to get 4 solutions
\beq 
\ve(\vk_\perp) = \gamma(\vk_\perp) \pm \sqrt{\hlts_y^2(\vk_\perp) + \left[ 1 + \hlts_z(\vk_\perp) \pm 1 \right]^2 }.
\eeq 
For the left edge state, we demand that 
\beq 
T(\ve_L(\vk_\perp), \vk_\perp) \vecenv{1}{0} = \lambda \vecenv{1}{0}; \quad |\lambda| < 1. 
\eeq 
Explicitly, this becomes 
\beq 
  \frac{1}{1 + \hlts_z} \vecenv{(\ve_L - \gamma)^2 - \Lambda^2}{\ve_L  - \gamma + \hlts_y} = \lambda \vecenv{1}{0}.
\eeq 
We can readily solve the the spectrum
\[
 \ve_L = \gamma - \hlts_y,
\]
while the decay condition becomes 
\[
 1 >  |\lambda| = \left| \frac{(\ve_L - \gamma)^2 - \Lambda^2}{1 + \hlts_z} \right| = |1 + \hlts_z|,
\]
which, using $\hlts_z > -2$, simply reduces to $\hlts_z < 0$. Similarly, for the right edge, we demand that 
\beq 
T(\ve_R(\vk_\perp), \vk_\perp) \vecenv{0}{1} = \lambda \vecenv{0}{1}; \quad |\lambda| > 1, 
\eeq 
which leads to
\[
 \ve_R = \gamma + \hlts_y, \quad \hlts_z < 0. 
\]

\bibliography{4node_wsm.bib}

\end{document}

%% file: defcommands.tex
\newcommand{\beq}{\begin{equation}}
\newcommand{\eeq}{\end{equation}}

\newcommand{\eq}[1]{eq. (\ref{#1})}

\newcommand{\tr}{\text{tr}}

\newcommand{\rank}{\text{rank}}
\newcommand{\sgn}[1]{\text{sgn}\left( #1 \right)}

\newcommand{\vecenv}[2]{\left( \begin{array}{c} #1 \\ #2 \end{array} \right)}

\newcommand{\cmplx}{\mathbb{C}}

\newcommand{\real}{\mathbb{R}}

\newcommand{\id}{\mathbbm{1}}

\newcommand{\SL}{\mathrm{SL}}

\newcommand{\hlt}{\mathcal{H}}

\newcommand{\green}{\mathcal{G}}

\newcommand{\bra}[1]{\langle #1 |}
\newcommand{\ket}[1]{| #1 \rangle}

\newcommand{\ve}{\varepsilon}

\newcommand{\vc}{\mathbf{c}}
\newcommand{\vcd}{\mathbf{c}^\dagger}

\newcommand{\vk}{\mathbf{k}}

\newcommand{\vv}{\mathbf{v}}
\newcommand{\vw}{\mathbf{w}}

\newcommand{\torus}{\mathbb{T}}